\documentclass[aps, prb, superscriptaddress, preprintnumbers]{revtex4}

\usepackage{graphicx}
\usepackage{amsmath}
\usepackage{amssymb}
\usepackage{color}
\usepackage{bm}

\newcommand{\be}{\begin{equation}}
\newcommand{\ee}{\end{equation}}
\newcommand{\ba}{\begin{eqnarray}}
\newcommand{\ea}{\end{eqnarray}}

\newcommand{\tr}{\,\mbox{tr}}
\newcommand{\sign}{\,\mbox{sgn}}

\definecolor{purple}{rgb}{0.8,0,0.6}

\begin{document}

\title{Engineering Weyl nodes in Dirac semimetals by a magnetic field}
\date{July 26, 2013}


\author{E. V. Gorbar}
\affiliation{Department of Physics, Taras Shevchenko National Kiev University, Kiev, 03022, Ukraine}
\affiliation{Bogolyubov Institute for Theoretical Physics, Kiev, 03680, Ukraine}

\author{V. A. Miransky}
\affiliation{Department of Applied Mathematics, Western University, London, Ontario, Canada N6A 5B7}

\author{I. A. Shovkovy}
\affiliation{School of Letters and Sciences, Arizona State University, Mesa, Arizona 85212, USA}
    
\begin{abstract}
We study the phase diagram of a Dirac semimetal in a magnetic field at a nonzero charge 
density. It is shown that there exists a critical value of the chemical potential at which a first-order phase 
transition takes place. At subcritical values of the chemical potential the ground state is a gapped state 
with a dynamically generated Dirac mass and a broken chiral symmetry. The supercritical phase is the 
normal (gapless) phase with a nontrivial chiral structure: it is a Weyl semimetal with a pair of 
Weyl nodes for each of the original Dirac points. The nodes are separated by a dynamically induced 
chiral shift. The direction of the chiral shift coincides with that of the magnetic field and its magnitude 
is determined by the quasiparticle charge density, the strength of the magnetic field, and the strength 
of the interaction. The rearrangement of the Fermi surface accompanying this phase transition is described.
\end{abstract}

\maketitle

\section{Introduction}
\label{1}

During past decades, a remarkable overlap of such seemingly different areas in physics as condensed 
matter and relativistic physics took place (for a recent review, see Ref.~\onlinecite{review}). It was 
especially clearly manifested in studying graphene.\cite{graphene} Such well known phenomena 
as the Klein paradox, the dynamics of a supercritical charge, and the dynamical generation of a 
Dirac mass in a magnetic field (magnetic catalysis), revealed in relativistic field theory, were first 
observed in studies of graphene (see Refs.~\onlinecite{Young, Wang, Nobel}). In the present paper, 
we will consider manifestations in condensed matter of another relativistic phenomenon: the 
rearrangement of the Fermi surface in three dimensional relativistic matter in a magnetic field. 
The original motivation for studying this phenomenon was its possible realization in magnetars 
and pulsars, and in heavy ion collisions.\cite{Gorbar:2009bm,Gorbar:2011ya} But as we discuss 
below, it could also be relevant for such new materials as Dirac and Weyl semimetals.       

Dirac and Weyl semimetals possess low-energy quasiparticles near the Fermi surface, which are 
described by the Dirac and Weyl equation, respectively.\cite{review} As was established long ago, 
an example of a semimetal whose low-energy effective theory includes three dimensional Dirac 
fermions is yielded by bismuth (for reviews, see Refs.~\onlinecite{Falkovsky,Edelman}). On the 
other hand, examples of the realization of the Weyl semimetals have been considered only 
recently.\cite{Wan,Burkov1,Burkov2} Weyl semimetals, which are three dimensional analogs of 
graphene, present a new class of materials with nontrivial topological properties.\cite{Volovik} 
Since their electronic states in the vicinity of Weyl nodes have a definite chirality, this leads to 
quite unique transport and electromagnetic properties of these materials.

The most interesting signatures of Dirac and Weyl semimetals discussed in the literature 
\cite{Burkov1,1210.6352,Franz,Carbotte,Basar,Abanin,Landsteiner} are connected with different nondissipative 
transport phenomena intimately related to the axial anomaly.\cite{anomaly} Many of them 
were previously suggested in studies of heavy ion collisions (for a review, see Ref.~\onlinecite{Notes}). 
 
In this paper, we will consider a different signature of Dirac semimetals: a dynamical 
rearrangement of their Fermi surfaces in a magnetic field. As we show 
below, this rearrangement is quite spectacular: a Dirac 
semimetal is transformed into a Weyl one. The resulting Weyl semimetal has a pair of 
Weyl nodes for each of the original Dirac points. Each pair of the nodes is separated by 
a dynamically induced (axial) vector $2\mathbf{b}$, whose direction coincides with the 
direction of the magnetic field. The magnitude of the vector $\mathbf{b}$ is determined 
by the quasiparticle charge density, the strength of the magnetic field, and the strength 
of the interaction. This phenomenon of the dynamical transformation of Dirac into Weyl 
semimetals is a condensed matter analog of the previously studied dynamical generation 
of the chiral shift parameter in magnetized relativistic matter in 
Refs.~\onlinecite{Gorbar:2009bm,Gorbar:2011ya}.

This paper is organized  as follows. In Sec.~\ref{section2} we introduce the model and set up the notations. 
The gap equation for the fermion propagator in the model is derived in Sec.~\ref{section3}. We show that, at 
a nonzero charge density, a pair of Weyl nodes necessarily arises in the normal phase of a Dirac metal 
as soon as a magnetic field is turned on. In Sec.~\ref{section4} a perturbative solution of the gap equation
describing the normal phase of the model is analyzed. A nonperturbative solution with a dynamical gap
that spontaneously breaks the chiral symmetry is analyzed in Sec.~\ref{section5}. A phase transition between
the normal phase and the phase with chiral symmetry breaking is revealed and described. In Sec.~\ref{section6}
we compare the dynamics in Dirac semimetals and graphene. A deep connection of the normal phase of a
Dirac metal in a magnetic field with the quantum Hall state with the filling factor $\nu = 2$ in graphene
is pointed out. The discussion of the results and conclusions is given in Sec.~\ref{section7}.
For convenience, throughout this paper, we set $\hbar=1$.

\section{Model}
\label{section2}

As stated in the Introduction, the main goal of this paper is to show that a dynamical 
transformation of Dirac semimetals into Weyl ones can be achieved by applying an external 
magnetic field to the former. It is convenient, however, to start our discussion
from writing down the general form of the low-energy Hamiltonian for a Weyl semimetal, 
\begin{equation}
H^{\rm (W)}=H^{\rm (W)}_0+H_{\rm int},
\label{Hamiltonian-model-Weyl}
\end{equation}
where
\begin{equation}
H^{\rm (W)}_0=\int d^3r \left[\,v_F \psi^{\dagger} (\mathbf{r})\left( 
\begin{array}{cc} \bm{\sigma}\cdot(-i\bm{\nabla}-\mathbf{b})  & 0\\ 0 & 
-\bm{\sigma}\cdot(-i\bm{\nabla}+\mathbf{b}) \end{array} 
\right)\psi(\mathbf{r})-\mu_{0}\, \psi^{\dagger} (\mathbf{r})\psi(\mathbf{r})
\right]
\label{free-Hamiltonian}
\end{equation}
is the Hamiltonian of the free theory, which describes two Weyl nodes of opposite (as required by 
the Nielsen--Ninomiya theorem \cite{NN}) chirality separated by vector $2\mathbf{b}$ in momentum 
space. In the rest of this paper, following the terminology of Refs.~\onlinecite{Gorbar:2009bm,Gorbar:2011ya},  
we will call $\mathbf{b}$ the chiral shift parameter. There are two reasons for choosing this terminology. First, as 
Eq.~(\ref{free-Hamiltonian}) implies, vector $\mathbf{b}$ shifts the positions of Weyl nodes from the origin in the momentum space
and, secondly, the shift has opposite signs for fermions of different chiralities. The other notations are: 
$v_F$ is the Fermi velocity, $\mu_{0}$ is the chemical potential, and
$\bm{\sigma}=(\sigma_x,\sigma_y,\sigma_z)$ are Pauli matrices associated with the 
conduction-valence band degrees of freedom in a generic low-energy model.\cite{Burkov3} Based on the similarity of the latter
to the spin matrices in the relativistic Dirac equation, we will call them pseudospin matrices.

The interaction part of the Hamiltonian describes the Coulomb interaction, i.e., 
\begin{equation}
H_{\rm int} = \frac{1}{2}\int d^3rd^3r^{\prime}\,\psi^{\dagger}(\mathbf{r})\psi(\mathbf{r})U(\mathbf{r}-\mathbf{r}^{\prime})
\psi^{\dagger}(\mathbf{r}^{\prime})\psi(\mathbf{r}^{\prime}). 
\label{int-Hamiltonian}
\end{equation} 
In order to present our results in the most 
transparent way, in this study we will utilize a simpler model with a contact four-fermion interaction,
\begin{equation}
U(\mathbf{r}) = \frac{e^2}{\kappa |\mathbf{r}|} \rightarrow g\, \delta^3(\mathbf{r}),
\label{model-interaction}
\end{equation}
where $\kappa$ is a dielectric constant and $g$ is a dimensionful coupling constant. 
As we argue in Sec.~\ref{section7}, such a model 
interaction should at least be sufficient for a qualitative description of the effect of 
the dynamical generation of the chiral shift parameter by a magnetic field in Dirac semimetals.

Before proceeding further with the analysis, we find it very convenient to introduce 
the four-dimensional Dirac matrices in the chiral representation:
\begin{equation}
\gamma^0 = \left( \begin{array}{cc} 0 & -I\\ -I & 0 \end{array} \right),\qquad
\bm{\gamma} = \left( \begin{array}{cc} 0& \bm{\sigma} \\  - \bm{\sigma} & 0 \end{array} \right),
\label{Dirac-matrices}
\end{equation}
where $I$ is the two-dimensional unit matrix, and rewrite our model Hamiltonian in a relativistic form,
\begin{equation}
H^{\rm (W)} = \int d^3 r\, \bar{\psi} (\mathbf{r})\left[
-i v_F (\bm{\gamma}\cdot \bm{\nabla})-(\mathbf{b}\cdot \bm{\gamma})\gamma^5-\mu_{0}\gamma^0 
\right]\psi(\mathbf{r})
+\frac{g}{2}\int d^3r\,\rho(\mathbf{r})\rho(\mathbf{r}),
\label{free-Hamiltonian-Weyl-rel}
\end{equation}
where, by definition, $\bar{\psi} \equiv \psi^{\dagger}\gamma^0$ is the Dirac conjugate spinor field, 
$\rho(\mathbf{r})\equiv \bar{\psi}(\mathbf{r}) \gamma^0 \psi (\mathbf{r})$ is the charge density 
operator, and the matrix $\gamma^5$ is
\begin{equation}
\gamma^5 \equiv i\gamma^0\gamma^1\gamma^2\gamma^3 
= \left( \begin{array}{cc} I & 0\\ 0 & -I \end{array} \right),
\end{equation}
where, as is clear from the first term in the free Hamiltonian (\ref{free-Hamiltonian}), the 
eigenvalues of $\gamma^5$ correspond to the node degrees of freedom.

The low-energy Hamiltonian of a Dirac semimetal corresponds to a special case in 
Eq.~(\ref{free-Hamiltonian-Weyl-rel}) when $\mathbf{b}=0$, i.e.,
\begin{equation}
H^{\rm (D)} = H^{\rm (W)}\Big|_{\mathbf{b} = 0}.
\label{free-Hamiltonian-Dirac}
\end{equation}
Unlike Weyl semimetals, Dirac semimetals are invariant under time reversal symmetry. 
Note that, for the clarity of presentation, in this paper we consider only the chiral limit 
when the bare Dirac mass term $m_0\bar{\psi} (\mathbf{r})\psi(\mathbf{r})$ is absent. 

In the presence of an external magnetic field, one should replace 
$\bm{\nabla} \to \bm{\nabla}+ie\mathbf{A}/c$, where $\mathbf{A}$ is the vector potential 
and $c$ is the speed of light. Thus, the Hamiltonian
of the Dirac semimetal model in an external magnetic field has the following form:
\begin{equation}
H^{\rm (D)}_{\rm mag} = \int d^3 r\, \bar{\psi} (\mathbf{r})\left[
-i v_F \left( \bm{\gamma}\cdot(\bm{\nabla}+ie\mathbf{A}/c) \right)-\mu_{0}\gamma^0 
\right]\psi(\mathbf{r})
+\frac{g}{2}\int d^3r\,\rho(\mathbf{r})\rho(\mathbf{r}).
\label{Hamiltonian-Dirac-magnetic}
\end{equation}
Note that both this Hamiltonian and the Weyl semimetal Hamiltonian (\ref{free-Hamiltonian-Weyl-rel}) are 
invariant under the chiral $U(1)_{+}\times U(1)_{-}$ symmetry, where $+$ and $-$ correspond to the node
states with $+1$ and $-1$ eigenvalues of the $\gamma_5$ matrix, respectively. The currents connected 
with the $U(1)_{+}$ and $U(1)_{-}$ symmetries are anomalous. However, because these symmetries are 
Abelian, one can introduce conserved charges for them.\cite{Jackiw}

\section{Gap equation}
\label{section3}

In this section, we will derive the gap equation for the fermion propagator in the 
Dirac semimetal model (\ref{Hamiltonian-Dirac-magnetic}) and show that at a nonzero 
charge density, a nonzero $\mathbf{b}$ {\it necessarily} arises in the normal phase as 
soon as a magnetic field is turned on. 

In model (\ref{Hamiltonian-Dirac-magnetic}), we easily find the following free fermion 
propagator:
\begin{equation}
iS^{-1}(u,u^\prime) = \left[(i\partial_t+\mu_0)\gamma^0
-v_F(\bm{\pi}\cdot\bm{\gamma})\right]\delta^{4}(u- u^\prime),
\label{sinverse}
\end{equation}
where $u=(t,\mathbf{r})$ and $\bm{\pi} \equiv -i \bm{\nabla} + e\mathbf{A}/c$ 
is the canonical momentum. In the rest of the paper, we will choose the vector potential 
in the Landau gauge, $\mathbf{A}= (0, x B,0)$, where $B$ is a strength of the external 
magnetic field pointing in the $z$ direction.

An ansatz for the full fermion propagator can be written in the following form (we will see 
that this ansatz is consistent with the Schwinger--Dyson equation for the fermion propagator 
in the mean-field approximation):
\begin{equation}
iG^{-1}(u,u^\prime)= \Big[(i\partial_t+\mu )\gamma^0 - v_F(\bm{\pi}\cdot\bm{\gamma}) 
+\gamma^0(\bm{\tilde{\mu}}\cdot\bm{\gamma})\gamma^5
+v_F (\mathbf{b} \cdot \bm{\gamma})\gamma^5
-m\Big]\delta^{4}(u- u^\prime).
\label{ginverse}
\end{equation}
This propagator contains dynamical parameters $\tilde{\bm{\mu}}$, $\mathbf{b}$, and $m$ 
that are absent at tree level in Eq.~(\ref{sinverse}). Here $m$ plays the role of the dynamical 
Dirac mass and $\mathbf{b}$ is the chiral shift.\cite{Gorbar:2009bm,Gorbar:2011ya} 
By taking into account the Dirac structure of the $\tilde{\bm{\mu}}$ term, we see that it is 
related to the anomalous magnetic moment $\mu_{\rm an}$ (associated with the pseudospin) 
as follows: $\tilde{\bm{\mu}}\equiv \mu_{\rm an}\mathbf{B}$.
It should be also emphasized that the dynamical parameter $\mu$ in the full propagator may 
differ from its tree-level counterpart $\mu_0$ [see Eq.~(\ref{gap-mu-text}) below].

In order to determine the values of these dynamical parameters, we will use the 
Schwinger--Dyson (gap) equation for the fermion propagator in the mean-field 
approximation, i.e.,
\begin{equation}
iG^{-1}(u,u^\prime) = iS^{-1}(u,u^\prime) - g \left\{ 
\gamma^0 G(u,u) \gamma^0 - \gamma^0\, \mbox{tr}[\gamma^0G(u,u)]\right\}
\delta^{4}(u- u^\prime). 
\label{gap}
\end{equation}
The first term in the curly brackets describes the exchange (Fock) interaction and
the last term presents the direct (Hartree) interaction. 

Separating different Dirac structures in the gap equation, we arrive at the following set
of equations:
\begin{eqnarray}
\mu - \mu_0  &=& -\frac{3}{4}\, g \, \langle j^{0}\rangle  ,
\label{gap-mu-text}  \\
\mathbf{b} &=& \frac{g}{4v_F } \, \langle\mathbf{j}_5\rangle    ,
\label{gap-Delta-text} \\
m &=& - \frac{g}{4} \,  \langle \bar{\psi}\psi\rangle   ,
\label{gap-m-text}\\
\bm{\tilde{\mu}} &=& \frac{g}{4} \, \langle \bm{\Sigma} \rangle  .
\label{gap-tilde-mu-text}
\end{eqnarray}
The fermion charge density, the axial current density, the chiral condensate, and the anomalous magnetic moment 
condensate on the right-hand side of the above equations are determined through the full fermion propagator as follows:
\begin{eqnarray}
\langle j^{0}\rangle &\equiv  &-\tr\left[\gamma^0G(u,u)\right] , 
 \label{density-text}\\
\langle \mathbf{j}_5\rangle &\equiv  &-\tr\left[\bm{\gamma}\,\gamma^5G(u,u)\right],
\label{axial-current-text}\\
\langle \bar{\psi}\psi\rangle&\equiv  & -\tr\left[G(u,u)\right]  ,
\label{chi-condensate-text}\\
\langle \bm{\Sigma} \rangle &\equiv  &-\tr\left[ \gamma^0\bm{\gamma}\,\gamma^5G(u,u)\right] . 
 \label{pseudospin-condensate-text}
\end{eqnarray}
Note that the right-hand sides in Eqs.~(\ref{chi-condensate-text}) and (\ref{pseudospin-condensate-text})
differ from those in Eqs.~(\ref{density-text}) and  (\ref{axial-current-text}) by the inclusion of an 
additional $\gamma^0$ matrix inside the trace. Since, according to Eq.~(\ref{Dirac-matrices}), 
the $\gamma^0$ matrix mixes quasiparticle states {from different Weyl nodes, we conclude 
that while $\langle j^{0}\rangle$ and $\langle \mathbf{j}_5\rangle$ describe the charge density
and the axial current density, the chiral condensate $\langle \bar{\psi}\psi\rangle$ and the 
anomalous magnetic moment condensate $\langle \bm{\Sigma} \rangle$ describe internode 
coherent effects.

As is known,\cite{Vilenkin,Metlitski:2005pr,Newman} in the presence of a fermion charge density and 
a magnetic field, the axial current $\langle \mathbf{j}_5\rangle$ is generated even in the free theory. 
Therefore, according to Eq.~(\ref{gap-Delta-text}), the chiral shift $\mathbf{b}$ is induced already in the lowest 
order of the perturbation theory. As a result, a Dirac semimetal is {\it necessarily} transformed into a Weyl one, 
as soon as an external magnetic field is applied to the system (see also a discussion in Sec.~\ref{section4}).

In order to derive the propagator $G(u,u^\prime)$ in the Landau-level representation, we invert 
$G^{-1}(u,u^\prime)$ in Eq.~(\ref{ginverse}) by using the approach described in Appendix A of
Ref.~\onlinecite{Gorbar:2011ya}. For our purposes here, the expression for the propagator in the 
coincidence limit $u^\prime\to u$ is sufficient [cf. Eq. (A26) in Ref.~\onlinecite{Gorbar:2011ya}]:
\begin{equation}
G(u,u)= \frac{i}{2\pi l^2}\sum_{n=0}^{\infty} \int\frac{d\omega d k^{3}}{(2\pi)^2}
\frac{{\cal K}_{n}^{-}{\cal P}_{-}+{\cal K}_{n}^{+}{\cal P}_{+}\theta(n-1)}{U_n},
\label{full-propagator}
\end{equation}
where ${\cal P}_{\pm}\equiv \frac12 \left(1\pm i s_\perp \gamma^1\gamma^2\right)$ are the pseudospin 
projectors, $l=\sqrt{c/|eB|}$ is the magnetic length, and  $s_\perp=\sign (eB)$. Also, by definition, $\theta(n-1)=1$ for $n\geq 1$ 
and $\theta(n-1)=0$ for $n < 1$. The functions ${\cal K}_{n}^{\pm}$ and $U_n$ with $n\geq 0$ are given by
\begin{eqnarray}
{\cal K}_{n}^{\pm}&=& \left[\left(\omega+\mu \mp s_{\perp}v_F b\right)\gamma^0
\pm s_{\perp}\tilde\mu + m -v_Fk^{3}\gamma^3\right]\Big\{
(\omega+\mu)^2 + \tilde{\mu}^2 -  m^2 - (v_F b)^2 - (v_Fk^{3})^2 - 2nv^2_F|eB|/c \nonumber \\
&& \mp 2s_{\perp}\left[\tilde\mu (\omega+\mu)+v_F b m \right]\gamma^0
\pm 2s_{\perp} (\tilde\mu + v_F b\gamma^0)v_Fk^{3} \gamma^3\Big\}
\label{K_n^pm}
\end{eqnarray}
and
\begin{equation}
U_n =\left[(\omega+\mu)^2 + \tilde{\mu}^2 - m^2 - (v_F b)^2 - (v_Fk^{3})^2 -2nv^2_F|e B |/c\right]^2
- 4\left[\left(\tilde{\mu}\,(\omega+\mu) + v_F b m\right)^2
        +(v_Fk^{3})^2\left((v_F b)^2-\tilde{\mu}^2\right)\right],
\label{U_n}
\end{equation}
where we took into account that the only nonvanishing components of the axial vectors 
$\mathbf{b}$ and $\bm{\tilde{\mu}}$ are the longitudinal projections $b$ and $\tilde{\mu}$ on the
direction of the magnetic field. Note that the zeros of the function $U_n$ determine the dispersion 
relations of quasiparticles.

\section{Perturbative solution}
\label{section4}

In order to obtain the leading order perturbative solution to the gap equations, we can use the free propagator 
on the right-hand side of Eqs.~(\ref{density-text}) through (\ref{pseudospin-condensate-text}), i.e.,
\begin{equation}
S(u,u)= \frac{i}{2\pi l^2}\sum_{n=0}^{\infty} \int\frac{d\omega d k^{3}}{(2\pi)^2}
\frac{ \left[\left(\omega+\mu_{0}\right)\gamma^0 -v_Fk^{3}\gamma^3\right]
\left[{\cal P}_{-}+{\cal P}_{+}\theta(n-1)\right]}{(\omega+\mu_{0})^2 - (v_F k^{3})^2 -2nv^2_F|eB|/c}.
\label{full-free-propagator}
\end{equation}
Note that unlike the high Landau levels with $n\geq 1$, where both spin projectors ${\cal P}_+$ 
and ${\cal P}_-$ contribute, the lowest Landau level (LLL) with $n=0$ contains only one projector ${\cal P}_-$.
The reason for this is well known. The Atiyah-Singer theorem connects the number of the zero energy 
modes (which are completely pseudospin polarized) of the two-dimensional part of the Dirac operator 
to the total flux of the magnetic field through the corresponding plane. This theorem states that  
LLL is topologically protected (for a discussion of the Atiyah-Singer theorem in the 
context of condensed matter physics, see Ref.~\onlinecite{Katsnelson}).

By making use of Eq.~(\ref{full-free-propagator}), we straightforwardly calculate the zeroth order result for the 
charge density,
\begin{equation}
\langle j^{0}\rangle_0=\frac{\mu_0}{2v_F(\pi l)^2}
+\frac{\sign(\mu_0)}{v_F(\pi l)^2}\sum_{n=1}^{\infty} \sqrt{\mu_0^2-2nv^2_F|eB|/c}\,\,
\theta\left(\mu_0^2-2nv^2_F|eB|/c\right),
\label{perturbative-charge}
\end{equation}
and the axial current density
\begin{equation}
\langle \mathbf{j}_5\rangle_0 = -\frac{e\mathbf{B}\mu_0}{2\pi^2 v_Fc}.
\label{perturbative-axial-current}
\end{equation}
As to the chiral condensate and the anomalous magnetic moment condensate, they vanish, 
i.e., $\langle \bar{\psi}\psi\rangle_0 = 0$ and $\langle\bm{\Sigma} \rangle_0 = 0$.
This is not surprising because both of them break the chiral $U(1)_{+} \times U(1)_{-}$ 
symmetry of the Dirac semimetal Hamiltonian (\ref{Hamiltonian-Dirac-magnetic}).
Then, taking into account Eqs.~(\ref{gap-m-text})} and (\ref{gap-tilde-mu-text}), 
we conclude that both the Dirac mass $m$ and the parameter $\tilde{\bm{\mu}}$ 
are zero in the perturbation theory.

After taking into account the gap equations (\ref{gap-mu-text}) and (\ref{gap-Delta-text}),
the results in Eqs.~(\ref{perturbative-charge}) and (\ref{perturbative-axial-current}) imply that     
there is a perturbative renormalization of the chemical potential and a dynamical generation 
of the chiral shift. Of special interest is the result for the axial current density given by 
Eq.~(\ref{perturbative-axial-current}). This is generated already in the free theory and 
known in the literature as the topological contribution.\cite{Son:2004tq,Metlitski:2005pr}
Its topological origin is related to the following fact: in the free theory, only the LLL contributes 
to both the axial current and the axial anomaly. By combining Eqs.~(\ref{gap-Delta-text}) and 
(\ref{perturbative-axial-current}), we find
\begin{equation}
\mathbf{b} = - \frac{ge\mathbf{B}\mu_0}{8\pi^2 v_F^2  c}.
\label{bvalue}
\end{equation}
This is our principal result, which reflects the simple fact that at $\mu_0 \neq 0$ (i.e., nonzero
charge density), the absence of the chiral shift is not protected by any additional symmetries 
in the normal phase of a Dirac metal in a magnetic field. Indeed, in the presence of a homogeneous 
magnetic field pointing in the $z$ direction, the rotational $SO(3)$ symmetry in the model is 
explicitly broken down to the $SO(2)$ symmetry of rotations around the $z$ axis. The 
dynamically generated chiral shift parameter $\mathbf{b}$ also points in the same direction 
and does  not break the leftover $SO(2)$ symmetry. The same is true for the discrete symmetries: 
while the parity ${\cal P}$ is preserved, all other discrete symmetries, charge conjugation 
${\cal C}$, time reversal ${\cal T}$, ${\cal CP}$, ${\cal CT}$, $PT$, and ${\cal CPT}$, are broken.
Last but not least, the chiral shift does not break the chiral $U(1)_{+}\times U(1)_{-}$ symmetry 
considered in Sec.~\ref{section2}. This implies that the dynamical chiral shift is necessarily generated 
in the normal phase of a Dirac semimetal in a magnetic field, and the latter is transformed into 
a Weyl semimetal.

\begin{figure}
\begin{center}
\includegraphics[width=.45\textwidth]{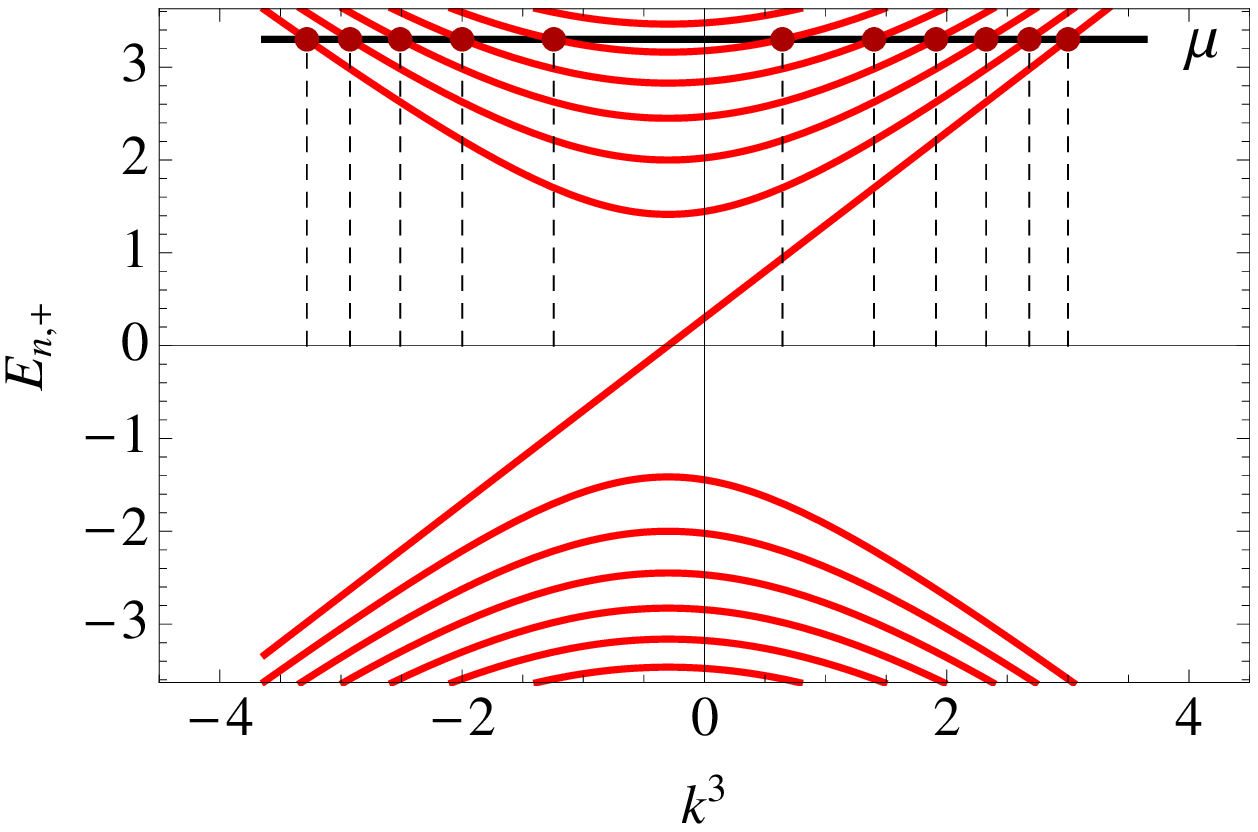}
\hspace{.04\textwidth}
\includegraphics[width=.45\textwidth]{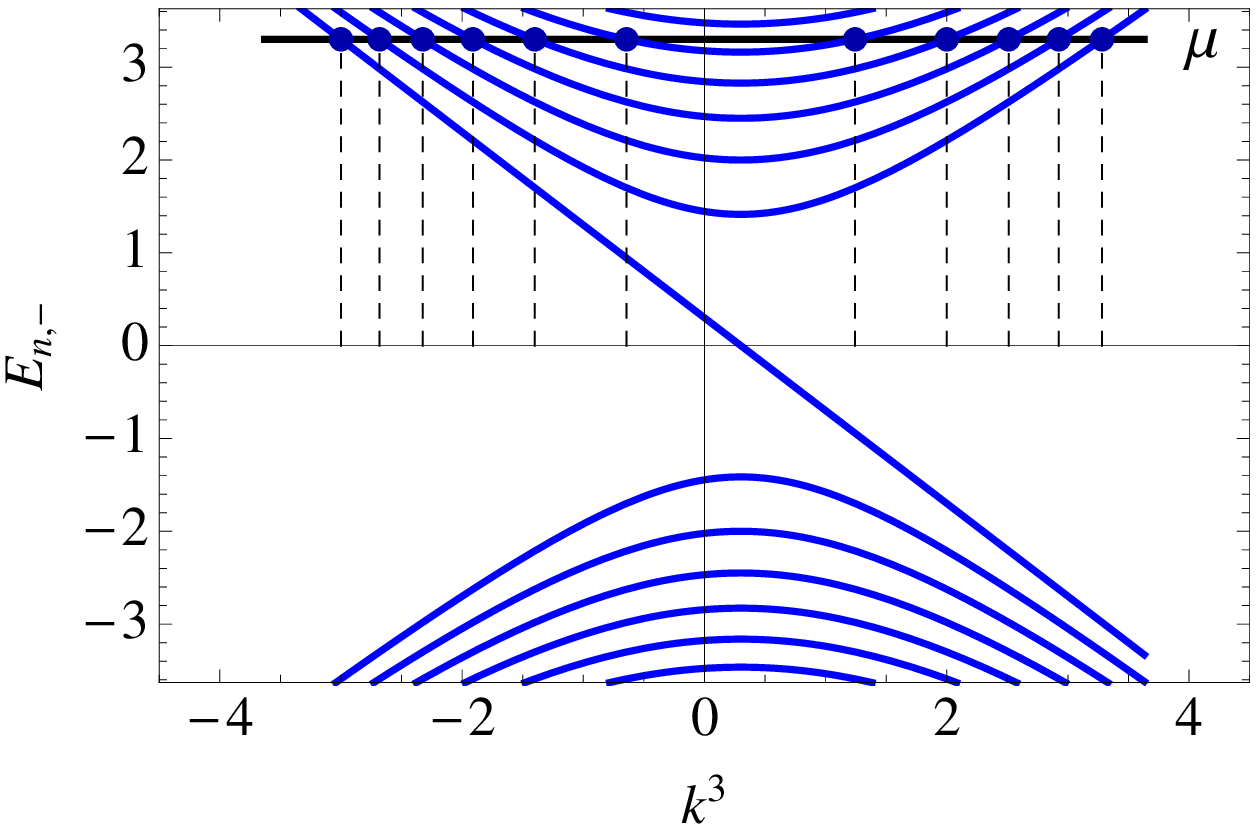}
\caption{Dispersion relations of the quasiparticles from different Weyl nodes and 
their Fermi surfaces. The two dispersion relations are the mirror images of 
each other.}
\label{fig:Fermi}
\end{center}
\end{figure}

In the phase with a dynamically generated chiral shift $\mathbf{b}$, the quasiparticle dispersion 
relations, i.e., $\omega_{n,\sigma} = -\mu + E_{n,\sigma}$, are determined by the following Landau-level 
energies (recall that we assume that the magnetic field points in the $z$ direction):
\begin{eqnarray}
E_{0, \sigma} &=& v_F \left( s_\perp b +\sigma k^3\right) ,\qquad n=0, \\
E_{n,\sigma} &=& \pm v_F \sqrt{\left(s_{\perp}b+\sigma k^3\right)^2+2n|eB|/c},\qquad n\geq 1,
\end{eqnarray}
where $\sigma=\pm$ corresponds to different Weyl nodes. The corresponding dispersion relations
are shown graphically in Fig.~\ref{fig:Fermi}. Note a qualitatively different character (compared to higher 
Landau levels) of the dispersion relations in the LLL given by the straight lines whose signs of slope 
correlate with the Weyl nodes. Of course, this correlation is due to the complete polarization of 
quasiparticle pseudospins in the LLL. As seen from Fig.~\ref{fig:Fermi}, the effect of the chiral shift
is not only to shift the relative position of the Weyl nodes in momentum space, but also to induce 
a chiral asymmetry of the Fermi surface.\cite{Gorbar:2009bm,Gorbar:2011ya} 

\section{Nonperturbative solution: Phase transition}
\label{section5}

The magnetic catalysis phenomenon, which takes place for planar as well as three-dimensional relativistic charged 
fermions because of the dimensional reduction,\cite{magCat,magCat3+1} implies that at vanishing 
$\mu_0$, the ground state in the model at hand is characterized by a nonzero Dirac mass $m$ that 
spontaneously breaks the chiral symmetry. (For the corresponding studies in graphene, see 
Ref.~\onlinecite{mc1}.)

Such a vacuum state can withstand a finite stress due to a nonzero chemical potential. However, as 
we discuss below, when $\mu_0$ exceeds a certain critical value $\mu_{\rm cr}$, the chiral symmetry 
restoration and a new ground state are expected. The new state is characterized by a nonvanishing 
chiral shift parameter $\mathbf{b}$ and a nonzero axial current in the direction of the magnetic field. 
Since no symmetry of the theory is broken, this state is the {\em normal} phase of the magnetized 
matter that happens to have a rather rich chiral structure. This phase was considered in the previous 
section.

Let us describe this transition in more detail. The value of the dynamical Dirac mass $m$ in the vacuum 
state can be easily calculated following the same approach as in Ref.~\onlinecite{magCat3+1}. At weak 
coupling, in particular, we can use the following expression for the chiral condensate: 
\begin{equation}
\langle \bar{\psi}\psi\rangle \simeq -\frac{m}{4\pi^2 v_F} 
\left(\Lambda^2 +\frac{1}{l^2}\ln\frac{v_F^2}{\pi m^2 l^2}  \right) ,
\label{condensate3+1}
\end{equation}
obtained in the limit of a small mass (which is consistent with the weak coupling approximation), using 
the gauge invariant proper-time regularization. Here the ultraviolet momentum cutoff $\Lambda$ can be related, for 
example, to the value of lattice spacing $a$ as follows: $\Lambda \simeq \pi/a$. Finally, by taking into 
account gap equation (\ref{gap-m-text}), we arrive at the solution for the dynamical mass,
\begin{equation}
m \simeq \frac{v_F}{\sqrt{\pi}l}\exp\left(-\frac{8 \pi^2 v_F l^2}{g}+\frac{(\Lambda l)^2}{2}\right).
\label{DiracMass}
\end{equation}
This zero-temperature, nonperturbative solution exists for $\mu_0< m$.

The free energies of the two types of states, i.e., the nonperturbative state with a dynamically 
generated Dirac mass (and no chiral shift) and the perturbative state with a nonzero chiral 
shift (and no Dirac mass) become equal at about $\mu_0\simeq m/\sqrt{2}$. This is analogous 
to the Clogston relation in superconductivity.\cite{Clogston}

At the critical value $\mu_{\rm cr} \simeq m/\sqrt{2}$, a first order phase transition takes place.
Indeed, both these solutions coexist at $\mu_0< m$, and while for $\mu_0 < \mu_{\rm cr}$
the nonperturbative (gapped) phase with a chiral condensate is more stable, the normal 
(gapless) phase becomes more stable at  $\mu_0 > \mu_{\rm cr}$. Note that at 
$\mu_0 < m$ the chemical potential is irrelevant in the gapped phase: the charge 
density is absent there. On the other hand, at any nonzero chemical potential, there is a 
nonzero charge density in the normal (gapless) phase. Therefore, at 
$\mu_{\rm cr} \simeq m/\sqrt{2}$, there is a phase transition with a jump in the charge 
density, which is a clear manifestation of a first order phase transition.

\section{Dirac semimetals vs. graphene in a magnetic field}
\label{section6}

It is instructive to compare the states in the magnetized Dirac semimetals and graphene. 
First of all, we would like to point out that the chiral shift is a three-dimensional
analog of the Haldane mass,\cite{Haldane,NSR} which plays an important role in the 
dynamics of the quantum Hall effect in graphene. Indeed, in the formalism of the 
four-component Dirac fields in graphene, the Haldane mass condensate is described 
by the same vacuum expectation value as that of the axial current in three 
dimensions:\cite{Gorbar:2008hu}
\begin{equation}
\langle \bar{\psi}\gamma^{3}\gamma^{5}\psi\rangle =  -\tr\left[\gamma^3\,\gamma^5G(u,u)\right]
\end{equation}
for a magnetic field pointing in the $z$ direction, which is orthogonal to the graphene plane;
cf. Eq.~(\ref{axial-current-text}). Moreover, similar to the solution with the chiral shift,
the solution with the Haldane mass (with the same sign for both spin-up and 
spin-down quasiparticles) describes the normal phase: it is a singlet with respect to the 
$SU(4)$ symmetry, which is a graphene analog of the chiral group in Dirac and Weyl 
semimetals. 

Also, in graphene, there is a phase transition similar to that described in the previous section. 
It happens when the LLL is completely filled.\cite{Gorbar:2008hu} In other words, the quantum 
Hall state with the filling factor $\nu = 2$ in graphene is associated with the normal phase 
containing the Haldane mass. 

As is well known, the Haldane mass leads to the Chern-Simons term in an external electromagnetic 
field.\cite{Haldane,NSR} This feature reflects a topological nature of the state with the filling factor 
$\nu = 2$ in a graphene. As was recently shown in Ref.~\onlinecite{Grushin:2012mt}, the chiral shift term 
\begin{equation}
\bar{\psi}(\mathbf{b} \cdot \bm{\gamma})\gamma^5\psi
\end{equation}
leads to an induced Chern-Simons term of the form 
$\frac{1}{2}b_\mu\epsilon^{\mu\nu\rho\sigma}F_{\rho\sigma}A_\nu$ in Weyl semimetals 
[here $b_\mu$ is a four-dimensional vector $(0,\mathbf{b})$]. Therefore it should also be 
generated in Dirac semimetals in a magnetic field. Note, however, the following principle 
difference between them: while in Weyl semimetals the chiral shift $\mathbf{b}$ is present  
in the free Hamiltonian, it is dynamically generated in the normal phase of Dirac semimetals 
in a magnetic field [see Eq.~(\ref{bvalue})]. Like in graphene, the generation of the 
Chern-Simons term implies a topological nature of the normal state in this material.\cite{footnote2}

\section{Discussion}
\label{section7}

In the present paper, we considered manifestations of a relativistic phenomenon, the 
rearrangement of the Fermi surface in three-dimensional matter in a magnetic field, 
in a Dirac semimetal. It was shown that its normal phase at a nonzero charge density 
and in a magnetic field has a nontrivial chiral structure: it is a Weyl semimetal with a 
pair of Weyl nodes for each of the original Dirac points. The nodes are separated by a 
dynamically induced chiral shift that is directed along the magnetic field. The phase 
transition between the normal phase and the phase with chiral symmetry breaking is revealed, 
and the rearrangement of the Fermi surface accompanying this phase transition is 
described.\cite{SekineNomura}

Although we studied a simple model with a contact four-fermion interaction, we believe that 
the present qualitative results apply equally well to more realistic models. The studies in the 
relativistic Nambu-Jona-Lasinio (NJL) model (with a contact interaction) on the one hand 
\cite{Gorbar:2009bm,Gorbar:2011ya} and in QED on the other \cite{Gorbar:2013upa} 
strongly support the validity of this statement. Namely, the dynamical generation of the 
chiral shift in a magnetic field and at a nonzero fermion density is a universal phenomenon. 

In the present study we analyzed the simplest model: it is isotropic (in the absence of 
the magnetic field), has no gap (i.e., no bare Dirac mass), and no Zeeman term for spin. 
In real materials, such as bismuth,\cite{Falkovsky,Edelman} the presence of an anisotropy, 
a gap, and the Zeeman term for spin should be taken into account. This task, as well as the 
generalization of the study to the case of nonzero temperature, is outside the scope of the 
present paper and will be considered elsewhere. Here we just want to mention that the 
effects of a bare Dirac mass term and temperature were studied in the NJL model in 
Ref.~\onlinecite{Gorbar:2011ya}. It was shown there that the chirality remains a good 
approximate quantum number even for massive fermions in the vicinity of the Fermi 
surface, provided the mass is sufficiently small, i.e., $m\ll\mu$. As for the temperature 
effects,\cite{Gorbar:2011ya} an interesting feature of the chiral shift is that it is insensitive 
to the temperature when $T \ll \mu$, and {\it increases} with temperature $T \gtrsim \mu$.

A natural extension of this work would be the study of the phase diagram of a Weyl semimetal 
in a magnetic field and a nonzero charge density. Because of a built-in chiral shift 
in the Hamiltonian of such a material, one may expect that its phase with chiral symmetry 
breaking, realized at zero (or low) density, should be inhomogeneous, possibly, a 
Larkin-Ovchinnikov-Fulde-Ferrel (LOFF) like one (cf. Refs.~\onlinecite{Ran,Chao,Zhang-chiral}).
We expect that a first order phase transition between the normal phase and the phase with
chiral symmetry breaking will take place in that case too.    

{\em Note added in proof.} Recently, an experimental observation
of the transition from a Dirac semimetal to a Weyl semimetal
in a magnetic field was reported.\cite{1307.6990}

\acknowledgments

We thank V.~P.~Gusynin for useful remarks. The work of E.V.G. was supported partially by the 
European FP7 program, Grant No.~SIMTECH 246937, SFFR of Ukraine, Grant No.~F53.2/028, 
and Grant STCU No.~5716-2 ``Development of Graphene Technologies and Investigation of 
Graphene-based Nanostructures for Nanoelectronics and Optoelectronics". The work of 
V.A.M. was supported by the Natural Sciences and Engineering Research Council of Canada. 
The work of I.A.S. was supported in part by the U.S. National Science Foundation 
under Grant No.~PHY-0969844.

\end{document}